\documentclass[prd,twocolumn,aps,superscriptaddress,nofootinbib,tightenlines]{revtex4}

\usepackage{amssymb}
\usepackage{braket}
\usepackage{slashed}

\def\nn{ \nonumber \\ }
\def\abs#1{\left| #1 \right|}
\def\vev#1{\left\langle #1 \right\rangle}

\def\pmns{\mathcal{U}}

\begin{document}

\title{Tribimaximal Mixing, Leptogenesis, and $\theta_{13}$}

\author{Elizabeth E.~Jenkins}

\author{Aneesh V.~Manohar}

\date{\today}

\begin{abstract}
We show that seesaw models based on flavor symmetries (such as $A_4$ and $Z_7 \rtimes Z_3$) which produce exact tribimaximal neutrino mixing also imply a vanishing leptogenesis asymmetry. We show that higher order symmetry breaking corrections in these models can give a non-zero leptogenesis asymmetry and generically also give deviations from tribimaximal mixing and a non-zero $\theta_{13} \agt 10^{-2}$.\\
\end{abstract}

\maketitle

Experiments using solar, atmospheric and reactor neutrinos, and neutrino beams produced at accelerators have confirmed the existence of neutrino  oscillations. The results are consistent with neutrino mixing produced if the neutrino weak eigenstates $\nu_e$,  $\nu_\mu$ and $\nu_\tau$ are related to the mass eigenstates $\nu_1$, $\nu_2$ and $\nu_3$ by a $3\times 3$ unitary matrix $\pmns$, commonly called the PMNS matrix,
\begin{eqnarray}
\ket{\nu_\alpha} &=& \pmns_{\alpha i} \ket{\nu_i}
\end{eqnarray}
where $\alpha \in \left\{e,\mu,\tau \right\}$ and $i \in \left\{1,2,3 \right\}$. The matrix $\pmns$ is written in terms of three angles $\theta_{12}$, $\theta_{13}$, and $\theta_{23}$, and three $CP$-violating phases $\delta$, $\alpha_1$ and $\alpha_2$~\cite{PDG},
\begin{eqnarray}
\pmns &=& \left[ \begin{array}{ccc} 
1 & 0 & 0 \\
0 & c_{23} & s_{23} \\
0 & -s_{23} & c_{23} \end{array} \right] \times
\left[ \begin{array}{ccc} 
c_{13} & 0 & s_{13}e^{-i \delta} \\
0 & 1 & 0 \\
- s_{13}e^{i \delta} & 0 & c_{13} \end{array} \right]\nn
&& \times  \left[ \begin{array}{ccc} 
c_{12} & s_{12} & 0 \\
-s_{12} & c_{12} & 0\\
0 & 0 & 1 \end{array} \right]\times
 \left[ \begin{array}{ccc} 
e^{i \alpha_1/2} & 0 & 0 \\
0 &e^{i \alpha_2/2} &0 \\
0 & 0 & 1 \end{array} \right]
\label{udef}
\end{eqnarray}
with $c_{ij} \equiv \cos \theta_{ij}$, $s_{ij} \equiv \sin \theta_{ij}$, and $0 \le \theta_{ij} \le \pi/2$, $0\le \delta,\alpha_{1,2} < 2\pi$. The Majorana phases $\alpha_{1,2}$ enter in lepton number violating amplitudes, and so are not observable presently in neutrino oscillation experiments, which measure lepton number conserving processes. The current experimental values of measured neutrino oscillation observables (taken from Ref.~\cite{review}) are:
\begin{eqnarray}
\Delta m_{21}^2 &=& \left(8.0 \pm 0.3 \right) \times 10^{-5} \ \text{eV}^2 \nn
\abs{\Delta m_{32}^2} &=& \left(2.5 \pm 0.2 \right) \times 10^{-3} \ \text{eV}^2 \nn
\tan^2 \theta_{12} &=& 0.45 \pm 0.05 \qquad \left( 30^\circ < \theta_{12} < 38^\circ\right)
\nn
\sin^2 2\theta_{23} &=& 1.02 \pm 0.04 \qquad \left( 36^\circ < \theta_{23} < 54^\circ\right)
\nn
\sin^2 2\theta_{13} &=& 0.0 \pm 0.05 \qquad \left(  \theta_{13} < 10^\circ\right).
\label{values}
\end{eqnarray}
There is an ongoing experimental program to measure or place an upper bound on $\theta_{13}$ at the level of
$\sin^2 2\theta_{13}  \sim 0.01$~\cite{dayabay}.

The ratio of the solar and atmospheric mass squared differences is $r=\Delta m_{21}^2/\abs{\Delta m_{32}^2}=(3.2 \pm 0.3) \times 10^{-2}$. Although the individual neutrino masses $m_i$ are not determined, the neutrino masses are known to be much smaller than the masses of all other standard model fermions from tritium endpoint, neutrinoless double beta decay and cosmological data. The smallness of neutrino masses can be naturally explained using the seesaw model~\cite{seesaw}, which extends the standard model by adding gauge singlet neutrinos.  The singlet neutrinos $N_R$ of the seesaw model naturally have Majorana masses much larger than the electroweak scale, unlike the standard model fermions which acquire mass proportional to electroweak symmetry breaking. An interesting feature of the seesaw model is that $CP$-violating decays of heavy singlet neutrinos can produce a lepton asymmetry in the early universe, which is converted into a baryon asymmetry at the electroweak scale. This leptogenesis mechanism~\cite{leptogenesis1,leptogenesis2} provides a simple explanation for the observed baryon asymmetry of the universe.

The neutrino mixing matrix has two large angles ($\theta_{12}$, $\theta_{23}$), and one small angle ($\theta_{13}$). A particularly interesting ansatz for the mixing matrix is the tribimaximal matrix~\cite{hps}
\begin{eqnarray}
\pmns_{TB}&=&\left[ \begin{array}{ccc} 
\sqrt{\frac{2}{3}} &\frac{1}{\sqrt3} &0\\[5pt]
-\frac{1}{\sqrt6} &\frac{1}{\sqrt3} &-\frac{1}{\sqrt2} \\[5pt]
-\frac{1}{\sqrt6} &\frac{1}{\sqrt3} &\frac{1}{\sqrt2} \\
\end{array}\right]  
\label{utb}
\end{eqnarray}
with $\tan^2\theta_{12}=1/2$, $\sin 2 
\theta_{23}=1$ and $\theta_{13}=0$.  The phase $\delta$ is undefined since $\theta_{13}=0$.
Eq.~\ref{utb} can be easily extended to include non-vanishing Majorana phases $\alpha_{1,2}$, $\pmns_{TB} \rightarrow \pmns_{TB} \ \text{diag}(e^{i \alpha_1/2}, e^{i \alpha_2/2}, 1 )$, which is the generalized form of tribimaximal mixing that we will consider in this work. The tribmaximal mixing matrix has been derived using models with discrete flavor symmetries. The models rely on the observation due to Ma~\cite{ma} that a Majorana mass matrix of the form
\begin{eqnarray}
\left[ \begin{array}{ccc} 
A & B & B \\ B & C & D \\ B & D & C \end{array}\right]  
\label{tbform}
\end{eqnarray}
is diagonalized by a mixing matrix with $\theta_{13}=0$ and $\sin^22\theta_{23}=1$. If $A+B=C+D$, then $\tan^2\theta_{12}=1/2$ and the mixing matrix is tribimaximal. The mixing matrix can have Majorana phases $\alpha_{1,2}$ if $A,B,C,D$ are complex. Particularly interesting are models based on the symmetries $A_4$~\cite{ma,altarelli} and $Z_7 \rtimes Z_3$~\cite{luhn}. These groups have a three-dimensional irreducible representation, and three inequivalent one-dimensional representations, so that the three generations of lepton doublets, charged leptons and singlet neutrinos can either transform as a  $\mathbf{3}$, or as three inequivalent one-dimensional representations, which distinguish between the generations.

It turns out that the seesaw models in the literature which derive exact tribimaximal mixing from a flavor symmetry do not allow for leptogenesis.  In these models, the low-energy neutrino mass matrix generated by the seesaw mechanism has the Ma form, but the product of neutrino Yukawa coupling matrices $Y_\nu^\dagger Y_\nu$ relevant to leptogenesis is proportional to the unit matrix, so the leptogenesis asymmetry parameter $\epsilon$ vanishes. This is true even if one considers the more general possibility of flavored leptogenesis~\cite{nardi,abada}. The tribimaximal mixing models have complex parameters, and have $CP$ violation. The low-energy PMNS matrix has $CP$ violation through non-zero $\alpha_{1,2}$. The problem is that the symmetry breaking pattern which generates a tribimaximal PMNS mixing matrix does not allow for $CP$ violation in the particular quantity $Y_\nu^\dagger Y_\nu$ that is needed for leptogenesis. The models typically have higher order corrections from higher dimension operators which are second order in flavor symmetry breaking. If the small symmetry breaking parameter is $\eta \ll 1$, we show that the leptogenesis asymmetry is order $\eta^2$. One can obtain $\epsilon \sim 10^{-6}$, which is the typical value necessary to obtain an adequate baryon asymmetry~\cite{leptogenesis2}, with $\eta \sim 10^{-2}$. We show that the flavor symmetry breaking also leads to deviations from tribimaximal mixing at first order in $\eta$, so that $\theta_{13}$ is typically non-zero, and larger than $\eta \sim  10^{-2}$.

Before proceeding further, we first review the standard seesaw scenario for neutrino  mixing and leptogenesis. The lepton mass terms in the seesaw theory are (following Ref.~\cite{broncano}):
 \begin{eqnarray}
{\cal L} &=& - \bar L_i \phi \left( Y_E \right)_{ij} E_{Rj}- \bar L_i \tilde \phi\left( Y_\nu 
\right)_{ij} N_{Rj} \nn
&&-\frac12 N_{Ri} M_{ij} N_{Rj} + \text{h.c.}
\label{seesaw}
\end{eqnarray}
where $i,j$ are flavor indices, $L=(e_L, \nu_L)$ are the lepton doublets, $E_R$ are charged lepton $SU(2)_L$ singlets with non-vanishing hypercharge, $N_R$ are gauge-singlet  fermion fields, and $\phi$ is the Higgs doublet ($\tilde \phi_\alpha =  \epsilon_{\alpha \beta} \phi^*_\beta$) with vacuum expectation value $v/\sqrt{2}$. The charged lepton mass matrix is $\left(m_E\right)_{ij} =  \left( Y_E \right)_{ij} v/\sqrt{2}$ and the Dirac neutrino mass matrix is $\left(m_D\right)_{ij}  = \left( Y_\nu \right)_{ij} v/\sqrt{2}$. 

One can make arbitrary flavor redefinitions $L \to U_L^{-1} L$, $N_R \to U_N^{-1} N_R $, $E_R \to U_E^{-1} E$ in Eq.~(\ref{seesaw}), where $U_{L,N,E}$ are $3 \times 3$ unitary matrices, under which
\begin{eqnarray}
M &\to& U_N^T \ M \ U_N \nn
Y_\nu &\to& U_L^{-1} \ Y_\nu \ U_N\nn
Y_E &\to& U_L^{-1} \ Y_E \ U_E\,.
\label{6}
\end{eqnarray}
It is convenient to pick a basis in which $M$ and $Y_E$ are diagonal, real, and non-negative, $M=\text{diag}(M_1,M_2,M_3)$, $Y_E=\text{diag}(y_e,y_\mu,y_\tau)$, which 
fixes $U_{L,N,E}$ up to a diagonal rephasing $U_L = U_E = \text{diag}(e^{i \zeta_1}, e^{i \zeta_2}, e^{i \zeta_3})$ which leaves $Y_E$ invariant.  In this basis, the only freedom to redefine $Y_\nu$ is given by a diagonal $U_L$ rephasing.  This rephasing can be used to eliminate three phases in $Y_\nu$, so the $3 \times 3$ complex matrix $Y_\nu$ contains 9 real and 6 imaginary physical parameters~\cite{broncano}.

The singlet Majorana mass matrix $M$ is not proportional to the weak scale $v$, and is naturally much larger than $v$ in unified theories. The Lagrangian Eq.~(\ref{seesaw}) leads to 
three heavy neutrinos with masses $M_i$ which are dominantly $N_R$, and three light neutrinos with masses of order $v^2/M$ which are dominantly $\nu_L$. Integrating out the 
heavy right-handed neutrinos leads to the dimension-five operator in the effective theory below $M$,
 \begin{eqnarray}
{\cal L}^{d=5} &=& \frac12 \left(\tilde \phi^\dagger L_i\right) \left(c_5 \right)_{ij} \left(\tilde \phi^\dagger 
L_j\right) 
+\text{h.c.}\label{lc5}
\end{eqnarray}
with
\begin{eqnarray}
c_5 &=&  Y_\nu^* {M^{*}}^{-1} Y_\nu^\dagger\,.
\label{c5}
\end{eqnarray}
When the Higgs field gets a vacuum expecation value, this generates a Majorana mass matrix $m=-(v^2/2)c_5$ for the light neutrinos. By definition, the PMNS matrix $\pmns$ diagonalizes $m \propto c_5$ in the basis in which $Y_E$ is diagonal,\footnote{The matrix $c_5$ is independent of the basis chosen for $N_R$, i.e.\ it is invariant under $U_N$ transformations.}
\begin{eqnarray}
-\frac{v^2}{2}\ \pmns^T c_5 \ \pmns &=& \text{diag}(m_1,m_2,m_3),
\end{eqnarray} 
where the light neutrino masses $m_i$ are real and non-negative. The masses $m_i$ and the PMNS matrix $\pmns$ are sufficient to describe neutrino physics at energies below $M$, and are the observables accessible in low-energy neutrino experiments. 

The PMNS  matrix, however, does not give complete information about the mixing structure of the seesaw Lagrangian.  The general form of $Y_\nu$ consistent with $\pmns$ and $m_i$ in a basis where $Y_E$ and $M$ are diagonal is~\cite{casasibarra}
\begin{eqnarray}
\frac{v}{\sqrt 2}\ Y_\nu &=& \pmns\, m^{1/2}\, O\, M^{1/2}
\label{12}
\end{eqnarray}
where $m=\text{diag}(m_1,m_2,m_3)$, $M=\text{diag}(M_1,M_2,M_3)$, and $O$ is a \emph{complex} orthogonal matrix $O O^T=1$. Note that in general $O$ is \emph{not} unitary. Low-energy physics fixes $\pmns$ and $m$, but leaves $O$ and $M$ undetermined.

At dimension six~\cite{broncano}, there is flavor mixing in the light neutrino kinetic terms after electroweak spontaneous symmetry breaking,
\begin{eqnarray}
{\cal L}^{d=6} &=& \left(c_6\right)_{ij} \left( \bar L_i \tilde \phi\right) i \slashed{\partial} \left( \tilde \phi^\dagger L_j\right)
\label{lc6}
\end{eqnarray}
with
\begin{eqnarray}
c_6 &=&  Y_\nu \left(M^\dagger M\right)^{-1} Y_\nu^\dagger.
\label{c6}
\end{eqnarray}
Measuring $c_6$ in addition to $\pmns$ and $m$ completely determines the parameters of the high-energy seesaw theory if the number of generations of heavy neutrinos is equal to the number of generations of standard model fermions.

Out-of-equilibrium decays in the early universe of $N_{Ri}$ to lepton and Higgs doublets produces lepton asymmetries.  In a basis where $M$ is diagonal and real, the lepton asymmetry parameters are~\cite{leptogenesis1,leptogenesis2,plumacher,covi}
\begin{eqnarray}
\epsilon_i &=& \frac{1}{8\pi \left(Y_\nu^\dagger Y_\nu\right)_{ii}}
\sum_{j\not=i} \text{Im}\left\{\left[\left( Y^\dagger_\nu Y_\nu\right)_{ij}\right]^2\right\}
f\left(\frac{\abs{M_j}^2}{\abs{M_i}^2}\right)\nn
\label{lepto}
\end{eqnarray}
where
\begin{eqnarray}
f(x) &=& \sqrt{x}\left[ \frac{2-x}{1-x}-(1+x)\ln \frac{1+x}{x}\right]
\stackrel{x\to\infty}{\longrightarrow} -\frac{3}{2\sqrt x}.\nn
\end{eqnarray}
For almost degenerate neutrinos,
\begin{eqnarray}
f(1+z) &\approx& -\frac{1}{z},\qquad z\ll1.
\label{deg}
\end{eqnarray}
The $\epsilon_i$ depend on the heavy neutrino masses, and the product
\begin{eqnarray}
Y_\nu^\dagger Y_\nu &=& \frac{2}{v^2}\ M^{1/2}\, O^\dagger\, m\, O\, M^{1/2},
\label{17}
\end{eqnarray}
using the form Eq.~(\ref{12}). One can have non-zero asymmetries $\epsilon_i$ if $O$ is complex. Note that $\pmns$ cancels out of $Y_\nu^\dagger Y_\nu$.

If leptogenesis takes place at temperatures below about $10^{12}$~GeV, then decoherence effects due to Yukawa interactions of the charged leptons are important, and the flavor of the charged leptons produced in the decays $N_{Ri} \to \ell_\alpha + \phi,\ \bar \ell_\alpha + \phi^\dagger$ are relevant. This scenario is referred to as flavored leptogenesis\footnote{We would like to thank E.~Nardi for helpful discussions on this point.}~\cite{nardi,abada}, and the lepton asymmetry depends on the asymmetry parameters~\cite{covi,plumacher,nardi,abada}
\begin{eqnarray}
\epsilon_i^{(\alpha)} &=& \frac{1}{8\pi \left(Y_\nu^\dagger Y_\nu\right)_{ii}}
\sum_{j\not=i}\nn
&& \text{Im}\biggl\{[\left(Y^\dagger_\nu\right)_{i\alpha}\left(Y_\nu\right)_{\alpha j}\left( Y^\dagger_\nu Y_\nu\right)_{ij}f(x_{j i}) \nn
&&+\left(Y^\dagger_\nu\right)_{i\alpha}\left(Y_\nu\right)_{\alpha j}\left( Y^\dagger_\nu Y_\nu\right)_{ji}\frac{1}{1-x_{j i}} \biggr\}\nn
x_{j i} &=& \frac{\abs{M_j}^2}{\abs{M_i}^2}
\label{flepto}
\end{eqnarray}
The flavor independent asymmetry parameter Eq.~(\ref{lepto}) is given by $\epsilon_i = \sum_{\alpha=e,\mu,\tau} \epsilon_i^{(\alpha)} $. $\pmns$ does not cancel in the combination  $\left(Y^\dagger_\nu\right)_{j\alpha}\left(Y_\nu\right)_{\alpha i}$ in Eq.~(\ref{flepto}).

There have been several studies of leptogenesis which assume that the PMNS matrix has tribimaximal form~\cite{Chan,Kitabayashi}.  It was shown that  mass matrices can be constructed in the seesaw Lagrangian which produce a large enough lepton asymmetry for the leptogenesis mechanism to lead to the baryon asymmetry of the universe.  However, this construction implicitly assumes that the tribimaximal form of the PMNS matrix is a low-energy accident, rather than a consequence of an underlying symmetry, as in the examples of Ref.~\cite{ma,altarelli,luhn}.

References~\cite{ma,altarelli,luhn} obtain the tribimaximal structure using a broken discrete flavor symmetry with a specific symmetry breaking pattern generated by the expectation values of scalar fields $\vev{\phi_i}$. The symmetry breaking structure that leads to tribimaximal mixing in these models gives no leptogenesis. If one includes additional symmetry breaking terms of higher order in $\vev{\phi}$, which exist via higher dimensional operators in the field theory, then one can avoid the leptogenesis problem. The higher dimension operators also lead to deviations from tribimaximal mixing. One can have non-zero leptogenesis while retaining the exact tribimaximal structure only if the higher dimension operators are tuned so that they do not perturb the PMNS matrix, in which case the exact tribimaximal form is accidental.

Another interesting class of models is based on $D_4$~\cite{grimus}, $S_3$~\cite{s3}, and $\mu\leftrightarrow \tau$ symmetry~\cite{mutau}. In these models, symmetry relates entries with $\mu \leftrightarrow \tau$, and the low-energy neutrino mass matrix has the Ma form Eq.~(\ref{tbform}) but without the restriction $A+B=C+D$, so that $\tan^2 \theta_{12}$ is not fixed to be $1/2$. These models typically give non-zero leptogenesis. However, if one wants the exact tribimaximal form with $\tan^2 \theta_{12}=1/2$ without any accidential fine tunings, then the leptogenesis asymmetry also vanishes. To be specific, in the $D_4$ model of Grimus and Lavoura~\cite{grimus},
\begin{eqnarray}
Y_\nu &=& \text{diag}(a,b,b)\nn
M &=& \left[ \begin{array}{ccc}M_1 & M_\chi & M_\chi \\
M_\chi & M_2 & 0 \\ M_\chi & 0 & M_2 \end{array}\right],
\end{eqnarray}
where all the parameters can be complex. This example has complex entries in $Y_\nu^\dagger Y_\nu$ in the basis in which $M$ is diagonal as long as $a\not=b$, but in general  $\tan^2 \theta_{12}\not=1/2$. Requiring that $c_5$ (Eq.~(\ref{c5})) have the Ma form Eq.~(\ref{tbform}) with $A+B=C+D$ for exact tribimaximal mixing leads to the constraint $a^2 M_2 = ab M_\chi + b^2 M_1$. To satisfy this relation requires an accidental fine-tuning between the Majorana mass matrix $M$ and the Dirac matrix $Y_\nu$, which are independent objects. One could obtain the constraint more naturally by assuming an additional symmetry of the underlying theory (as happens in the $A_4$ model) that restricts $Y_\nu$ and $M$ separately by $a=b$ and $M_2=M_\chi+M_1$. But then the leptogenesis asymmetry vanishes.

In the remainder of the paper, we will use the specific seesaw implementation of $A_4$ symmetry given by Altarelli and Feruglio~\cite{altarelli} to illustrate our point about the incompatibility of tribimaximal mixing derived from an exact flavor symmetry with leptogenesis. $A_4$ is the symmetry group of the tetrahedron, or the group of even permutations on four objects, and has order 12. It has three inequivalent one dimensional representations, $\mathbf{1}$, $\mathbf{1^\prime}$, $\mathbf{1^{\prime\prime}}$ and a three dimensional representation $\mathbf{3}$.

Altarelli and Feruglio use a supersymmetric theory with an $A_4 \otimes Z_3$ discrete flavor symmetry. In addition to the standard model and singlet neutrino multiplets, they have scalar fields $\phi_S$ and $\phi_T$ which transform as $A_4$ triplets, and $\xi$ which transforms as an $A_4$ singlet.  These scalars  develop vacuum expectation values, $\vev{\phi_T}=(v_T,0,0)$, $ \vev{\phi_S}=(v_S,v_S,v_S)$, and $\vev{\xi}=u$ which break $A_4 \times Z_3$. There also are additional fields  needed to construct a superpotential, which are not important for our analysis. The $A_4 \otimes Z_3$ representations of the relevant fields are given in Table~\ref{tab} along with their $U(1)_R$ charges. The  standard model Higgs multiplets $H_{u,d}$ are $A_4 \times Z_3$ singlets, with vacuum  expectation values $v_{u,d}$.
\begin{table}
\begin{eqnarray*}
\begin{array}{c|cccccccc}
 &e^+ & \mu^+ & \tau^+ & L_L & N^c_L & \phi_S & \phi_T & \xi \\
 \hline
A_4 & 1 & 1^{\prime\prime} & 1^\prime & 3 & 3 & 3 & 3 & 1\\
Z_3 &  \omega^2 & \omega^2 & \omega^2 & \omega & \omega^2 & \omega^2
& 1 & \omega^2\\
U(1)_R & 1 & 1 & 1 & 1 & 1 & 0 & 0 &0
\end{array}
\end{eqnarray*}
\caption{\label{tab} Transformation properties of the fields under $A_4 \otimes Z_3 \otimes U(1)_R$. }
\end{table}
The structure of the lepton mass matrices then follows from a standard spurion analysis, 
assuming higher dimension operators are suppressed by a high scale $\Lambda$. The 
expansion parameter is $\eta=V/\Lambda$, where we use $V \sim v_T,v_S,u$ to denote the 
typical $A_4\otimes Z_3$ flavor symmetry breaking expectation values of the scalar fields. 

We first summarize the seesaw model of Altarelli and Feruglio. The superpotential terms $ x_A \xi N^c N^c$, $x_B \phi_S N^c N^c$, and $y N^c L H_u$, where $x_A,x_B,y$ are coupling constants, generate the matrices
\begin{eqnarray}
M^\dagger &=& 2x_A u \left[\begin{array}{ccc} 
1 & 0 & 0 \\ 0 & 0 & 1 \\ 0 & 1 & 0\end{array}\right]+ \frac23 x_B v_S \left[\begin{array}
{ccc}
2 & -1 & -1 \\ -1 & 2 & -1 \\ -1 & -1 & 2 \end{array}\right]\nn
Y_\nu^\dagger &=& y  \left[\begin{array}{ccc} 
1 & 0 & 0 \\ 0 & 0 & 1 \\ 0 & 1 & 0\end{array}\right]\,.
\label{1}
\end{eqnarray}
The leading contribution to the charged lepton masses is from the superpotential terms $y_e e^+ \left(\phi_T L\right)_1 H_d/\Lambda$, $y_\mu \mu^+ \left(\phi_T L \right)_{1^{\prime}} H_d/\Lambda$, $y_\tau \tau^+ \left(\phi_T L\right)_{1^{\prime\prime}}  H_d/\Lambda$ which are higher dimension operators\footnote{$\left(\phi_T L\right)_{1,1^\prime,1^{\prime\prime}}$ denotes that $\phi_T$ and $L$ are combined to form the $A_4$ representations $1,1^\prime,1^{\prime\prime}$.} suppressed by one power of $
\Lambda$,
\begin{eqnarray}
Y_E^\dagger &=& \frac{v_T}{\Lambda}\left[\begin{array}{ccc}
y_e & 0 & 0 \\ 0 & y_\mu & 0 \\ 0 & 0 & y_\tau\end{array}\right]\,.
\label{2}
\end{eqnarray}
It is necessary to introduce higher dimension operators to get non-zero charged lepton masses. The Yukawa coupling $y_{\tau}$ is of order $\eta$, so $ \eta \agt 10^{-2}$ to get a large enough $\tau$ mass.

It is simple to verify that Eqs.~(\ref{1},\ref{2}) lead to tribimaximal mixing, with heavy neutrinos of mass $M_1 = \abs{2x_A u + 2x_B v_S} $, $M_2=\abs{2 x_A u} $, $M_3= \abs{-2 x_A u+2 x_B v_S}$, and light neutrinos of mass $m_i = \abs{y v_u}^2/M_i$~\cite{altarelli}.  Let $2\phi_{1,2,3}$ be the phases of $ 2x_A u + 2x_B v_S$, $2 x_A u$, $-2 x_A u+2 x_B v_S$, respectively, and $\phi_y$ be the phase of $y$.\footnote{The 
two-fold ambiguity in $\phi_{1,2,3}$ is irrelevant (see e.g.\ Ref.~\cite{rephasing}).} Then the PMNS matrix is $\pmns_{TB}e^{i\Phi}$ with $\Phi=\text{diag}(\phi_1-\phi_y,\phi_2-\phi_y, \phi_3-\phi_y)$, which can be converted by a phase redefinition into $\Phi=\text{diag} (\phi_1-\phi_3,\phi_2-\phi_3,0)$, which is the standard form with only two Majorana phases. Since the three $M_i$ are given in terms  of two complex numbers $2x_Au$ and $2x_B v_S$, it is not possible to have arbitrary values for $M_i$. For the case of normal hierarchy, $m_1 < m_2 < m_3$, and $\Delta m_{21}^2 \ll \abs{\Delta m_{32}^2}$, so that $2x_A u \approx 2 x_B v_S$ (which requires a fine-tuning at the level of $1/r \sim 30$ between the two terms), $M_1\approx 2M_2$, and $\phi_1 \approx \phi_2$. The known neutrino mass differences then imply that $M_3 \approx \sqrt{4 r/3}\,M_2 \approx M_2/5$. Equivalently, $m_1=m_2/2=  \sqrt{r/3}\, m_3$, so that $m_1=5.2 \times 10^{-3}$~eV, $m_2=1.03 \times 10^{-2}$~eV and $m_3=0.05$~eV.
For the inverted hierarchy, $M_1 \approx M_2 \approx M_3/3$ and $(M_1-M_2)=4rM_3/27$~\cite{altarelli}, so that $m_3 = 1.8 \times 10^{-2}$~eV, and $m_1 \approx m_2 =5.3 \times 10^{-2}$~eV.

It follows from Eq.~(\ref{1}) that 
\begin{eqnarray}
Y_\nu^\dagger Y_\nu = \abs{y^2} \mathbf{1},
\label{nolg}
\end{eqnarray}
so that there is no leptogenesis (see Eq.~(\ref{lepto})). There is also no flavored leptogenesis (see Eq.~(\ref{flepto})). The combinations
$\left(Y^\dagger_\nu\right)_{i\alpha}\left(Y_\nu\right)_{\alpha j}\left( Y^\dagger_\nu Y_\nu\right)_{ij}$
and $\left(Y^\dagger_\nu\right)_{i\alpha}\left(Y_\nu\right)_{\alpha j}\left( Y^\dagger_\nu Y_\nu\right)_{ji}$
in the flavored leptogenesis asymmetries $\epsilon_i^{(\alpha)}$ contain one factor of $Y^\dagger_\nu Y_\nu$ in which the charged lepton index has been summed over. Equation~(\ref{nolg}) implies that
$Y^\dagger_\nu Y_\nu$ is diagonal and proportional to the unit matrix;  thus $\epsilon_i^{(\alpha)}$ vanish since they contain a factor of the off-diagonal elements $i\not=j$ of $Y^\dagger_\nu Y_\nu$.

While we have used a specific $A_4$ model, this conclusion is common to all models in the literature which generate exact tribimaximal mixing from a symmetry.  (Note that in the case of the studied $A_4$ model, the assignment of three-dimensional irreducible representations to the singlet neutrinos and to the lepton doublets forces the $Y_\nu$ matrix to take the given form.)

There are corrections to Eq.~(\ref{1},\ref{2}) from operators suppressed by higher powers of $1/\Lambda$. The leading correction to $Y_\nu$ is from $x_C N^c \left(L \phi_T\right)_{3_S} H_u /\Lambda$ and $x_D N^c \left(L \phi_T\right)_{3_A} H_u /\Lambda$ where $3_{S,A}$ are the triplets in the symmetric and antisymmetric product $3 \otimes 3$, and gives
\begin{eqnarray}
\delta^{(1)} Y_\nu^\dagger &=& \frac{x_C v_T}{3 \Lambda}\left[\begin{array}{ccc}
2 & 0 & 0 \\ 0 & 0 & -1 \\ 0 & -1 & 0\end{array}\right]
+\frac{x_D v_T}{2 \Lambda}\left[\begin{array}{ccc}
0 & 0 & 0 \\ 0 & 0 & 1 \\ 0 & -1 & 0\end{array}\right]\,.\nn
\label{3}
\end{eqnarray}

The leading corrections to $M$ are from the terms $x_E \left(N^c N^c\right)_{3_S} \left( \xi \phi_T \right)/\Lambda$ and $x_{RR^\prime}\left(N^c N^c\right)_{R} \left(\phi_S \phi_T \right)_{R^\prime}/\Lambda$, where  $(R,R^\prime)=(1,1)$, $(1^\prime,1^{\prime\prime}) $, $(1^{\prime\prime},1^\prime)$, $(3_S,3_S)$, and $(3_S,3_A)$. They give a correction to $M$ of the form
\begin{eqnarray}
\delta^{(1)} M^\dagger &=& \frac{2 x_E u v_T}{3 \Lambda}\left[\begin{array}{ccc}
2 & 0 & 0 \\ 0 & 0 & -1 \\ 0 & -1 & 0\end{array}\right]
+\frac{x_{1^\prime 1^{\prime\prime}} v_S v_T}{2 \Lambda}\left[\begin{array}{ccc}
0 & 1 & 0 \\ 1 & 0 & 0 \\ 0 & 0 & 1\end{array}\right]\nn
&&+\frac{x_{1^{\prime\prime}1^\prime  } v_S v_T}{2 \Lambda}\left[\begin{array}{ccc}
0 & 0 & 1 \\ 0 & 1 & 0 \\ 1 & 0 & 0\end{array}\right]\,.
\label{4}
\end{eqnarray}
The $(3_S,3_S)$, $(3_S,3_A)$ and $(1,1)$ terms can be absorbed in redefinitions of $x_{A,B,E}$.

The leading corrections to $Y_E$ are from $\left(e^+ L\right)_{3} \left(\phi_T^2\right)_{3_S}  H_d/\Lambda^2$, $\left(\mu^+ L\right)_{3} \left(\phi_T^2\right)_{3_S} H_d/\Lambda^2$ and $\left(\tau^+ L\right)_{3} \left(\phi_T^2\right)_{3_S} H_d/\Lambda^2$. Since the vacuum expectation value of $\left(\phi_T^2\right)_{3_S}$ is proportional to $\vev{\phi_T}$, these corrections can be absorbed into a redefinition of $y_{e,\mu,\tau}$ in Eq.~(\ref{2}).

There can be direct higher dimension operator contributions to $c_5$, which are not generated by the seesaw mechanism~\cite{altarelli}. These arise from operators such as $LLH_uH_u \xi^2/\Lambda^3$, $LLH_uH_u \xi \phi_S/\Lambda^3 $, $LLH_uH_u \phi_S^2/\Lambda^3$, and produce neutrino mass terms of order $v_u^2 V^2/\Lambda^3$. They are of order $V^3/\Lambda^3$ relative to the seesaw generated mass terms of order $v_u^2/V$, and can be neglected in our analysis.

The corrections Eq.~(\ref{3},\ref{4}) lead to deviations from exact tribimaximal mixing, and from the diagonal form Eq.~(\ref{nolg}), which can be computed in perturbation theory. It is convenient to go to a basis where the lowest order matrices Eq.~(\ref{1},\ref{2}), which we now denote by $Y^\dagger_{\nu 0}$ and $M^\dagger_0$, are diagonal.  Let
\begin{eqnarray}
\widetilde M^\dagger &\equiv& \pmns_{0}^\dagger M^\dagger \pmns_{0}^*\nn
\widetilde Y_\nu^\dagger &\equiv& \pmns_0^\dagger Y_\nu^\dagger \pmns_0
\end{eqnarray}
with $\pmns_0=\pmns_{TB}e^{i\Phi_0}$, $\Phi_0=\text{diag}(\phi_1,\phi_2,\phi_3)$, and $\widetilde M = \widetilde M_0 + \delta \widetilde M$, where $\delta \widetilde M =\delta^{(1)} \widetilde M + \delta^{(2)}\widetilde M + \ldots$ is an expansion in powers of $V/\Lambda$. Let $S$ be the unitary matrix which makes $S^\dagger \widetilde M^\dagger S^*$ diagonal, real, and non-negative, and write $S=\exp(i s)$, where $s$ is hermitian, and has the expansion $s = \delta^{(1)}s + \delta^{(2)}s +\ldots$. In the basis where $M$ is diagonal, $Y_\nu^\dagger Y_\nu$ becomes $S^\dagger \widetilde Y_\nu^\dagger \widetilde Y_\nu S$, which is
\begin{eqnarray}
&& \abs{y}^2 \mathbf{1} +  \Bigl(\widetilde Y_{\nu 0}^\dagger\ \delta^{(1)} \widetilde Y_\nu 
+ \delta^{(1)} \widetilde Y_\nu^\dagger\ \widetilde Y_{\nu 0}\Bigr)+\ldots\nn
&=&  \abs{y}^2 \mathbf{1} + 
\left[\begin{array}{ccc}
1 & \sqrt{2} e^{i \phi_{21} } & 0 \\ \sqrt{2} e^{-i \phi_{21}} & 0 & 0 \\ 0 & 0 & -1\end{array}
\right]\text{Re}\left( \frac{2 x_C v_T y^*}{3 \Lambda}\right)\nn
&&\hspace{-0.7cm} + 
\left[\begin{array}{ccc}
0 & 0 & \sqrt{\frac13} e^{i \phi_{31}} \\ 0 & 0 & - \sqrt{\frac23} e^{i \phi_{32}} \\ 
\sqrt{\frac13} e^{-i \phi_{31}} & - \sqrt{\frac23} e^{-i \phi_{32}} & 0\end{array}\right]
\text{Re}\left( \frac{x_D v_T y^*}{ \Lambda}\right)\nn
\label{19}
\end{eqnarray}
up to first order in $\eta=V/\Lambda$, where $\phi_{ij}=\phi_i-\phi_j$. $S$ cancels since $Y_{\nu 0}^\dagger Y_{\nu 0} \propto \mathbf{1}$.    
It can be seen from Eq.~(\ref{19}) that $\text{Im}\{[( Y^\dagger_\nu Y_\nu)_{ij}]^2\}$ is non-zero and order $\eta^2$ in the basis where $M$ is diagonal.

For the normal hierarchy, the lightest right-handed neutrino is $M_3$, and leptogenesis is governed by $\epsilon_3$, i.e.\ by $\left(Y^\dagger_\nu Y_\nu\right)_{3i}$, $i=1,2$, which are non-zero, and complex. Hence Eq.~(\ref{19}) leads to non-zero leptogenesis for the case of normal hierarchy. Using Eq.~(\ref{lepto}), the lowest order values $M_1 \approx 2 M_2$, $\phi_1 \approx \phi_2$, and the asymptotic form for $f(x)$ for normal hierarchy gives
\begin{eqnarray}
\epsilon_3 &=& \frac5{32\pi } \frac{M_3}{M_2}\left[\text{Re}\left( \frac{x_D v_T e^{-i\phi_y}}{ \Lambda}\right)\right]^2\sin2 \phi_{32},
\label{20}
\end{eqnarray}
combining the $M_2$ and $M_1$ terms. For the inverted hierarchy, $M_2$ is the lightest right-handed neutrino, and $\epsilon_2$ controls leptogenesis,
\begin{eqnarray}
\epsilon_2 &=& -\frac1{8\pi} \frac{M_2}{M_3}\left[\text{Re}\left( \frac{x_D v_T y^*}{ \Lambda}\right)\right]^2\sin2 \phi_{32}\nn
&&+\frac{9 }{32 \pi r }\left[\text{Re}\left( \frac{2x_C v_T y^*}{ 3\Lambda}\right)\right]^2\sin2 \phi_{21}.
\label{21}
\end{eqnarray}
In the second term, we have used Eq.~(\ref{deg}) for $f(M_1^2/M_2^2)$. 

The PMNS matrix is $\pmns=\pmns_{TB}e^{i\Phi_0}  \pmns_1 $, where $\pmns_1$ is defined so that $\pmns_1^T \widetilde c_5 \pmns_1$ is diagonal, with $\widetilde c_5\equiv  e^{-2i\phi_y}
\widetilde Y_\nu^* \hbox{${\widetilde M}^*$}^{-1}\widetilde Y_\nu^\dagger$. Writing $\pmns_1=\exp{i X}$ with $X$ hermitian, and expanding $X=\delta^{(1)}X + \ldots $, we can solve for $X$ to first order in $V/\Lambda$,
\begin{eqnarray}
\delta^{(1)} X_{ii} &=& - \frac{\text{Im} \left[\delta^{(1)}\tilde c_5\right]_{ii}}{2 \left[c_{\tilde 
5,0}\right]_i} \nn
\delta^{(1)} X_{i\not=j} &=& - \frac{\text{Im} \left[\delta^{(1)}\tilde c_5\right]_{ij}}
{\left[c_{\tilde 5,0}\right]_i+\left[c_{\tilde 5,0}\right]_j}+i\frac{\text{Re} \left[\delta^{(1)}\tilde 
c_5\right]_{ij}}{\left[c_{\tilde 5,0}\right]_i-\left[c_{\tilde 5,0}\right]_j} \nn
\label{28}
\end{eqnarray}
where $\tilde c_5 = \tilde c_{5,0}+\delta^{(1)} \tilde c_5+\ldots$, and $\tilde c_{5,0}$ is diagonal. The eigenvalues are
\begin{eqnarray}
\Lambda_i &=& \left[\tilde c_{5,0}\right]_{ii} + \text{Re} \left[\delta^{(1)}\tilde c_{5,0}\right]_{ii}+\ldots
\end{eqnarray}
The light-neutrino mass shifts are $\delta m_i/m_i \sim \mathcal{O}\left(V/\Lambda\right)$ due to the higher dimension operators.

The elements of $X_{ij}$ are of order $V/\Lambda$, and lead to deviations from exact tribimaximal mixing. To first order,
\begin{eqnarray}
\abs{\pmns_{13}} &=& s_{13}\nn
&=& \frac{1}{\sqrt3}\abs{\sqrt{2}e^{i \phi_{13}} X_{13}+e^{i\phi_{23}}X_{23}}.
\end{eqnarray}
Using $\abs{{\pmns_{12}}/{\pmns_{11}}} = \tan \theta_{12}$, $\abs{{\pmns_{23}}/{\pmns_{33}}} = \tan \theta_{23}$, 
which are rephasing invariant expressions for the mixing angles~\cite{rephasing}, and letting $\theta_{12}=\tan^{-1}(1/\sqrt2)+\delta \theta_{12}$, $\theta_{23}=\pi/4+\delta \theta_{23}$ gives
\begin{eqnarray}
\delta \theta_{12}
&=& -\sqrt2\,\text{Im}\left[e^{i\phi_{12}}X_{12} \right]
\end{eqnarray}
\begin{eqnarray}
\delta \theta_{23}
&=&-\frac{1}{\sqrt{3}}\text{Im}\left[e^{i\phi_{13}}X_{13}-\sqrt2e^{i \phi_{23}}X_{23} \right]
\label{32}
\end{eqnarray}
Eqs.~(\ref{28})--(\ref{32}) are general, and give the first order correction to the PMNS matrix in terms of the correction to $c_5$.\footnote{Higher order corrections perturbing the tribimaximal form of the PMNS matrix have been studied in an RS model recently~\cite{csaki}.}

The requirement that the model produce a large enough lepton asymmetry gives $\epsilon \sim 10^{-6}$. From Eq.~(\ref{20},\ref{21}), and using $M_3/M_2 \sim 1/5$ for the normal hierarchy, and $M_2/M_3 \sim 1/3$ for the inverted hierarchy gives the estimates $\epsilon_3 \sim \eta^2/(32 \pi)$ and $\epsilon_2 \sim 9\eta^2/(32\pi r)$, respectively. We already have the lower bound $\eta \agt 10^{-2}$ from the charged lepton masses, so the model can produce an adequate lepton asymmetry ($ \epsilon \agt 10^{-6}$) when higher order terms in $\eta$ are included in the mass matrices.

The higher order terms also lead to deviations from tribimaximal mixing at order $\eta$, and in particular, $s_{13} \sim \sqrt{\epsilon} \sim \eta \agt 10^{-2}$. There are many parameters that enter the mass matrix at order $\eta$, and we have not done a complete analysis of the parameter space. As a simple example, consider the case where the only higher order coefficient that is non-zero is $x_D$ which enters into Eq.~(\ref{19}). The non-zero elements of $\delta c$ are:
\begin{eqnarray}
\delta c_{13}
&=&-\frac{2\sqrt3\abs{y}^2 x_B x_D}{M_1M_3}e^{-i\left(\phi_1+\phi_3\right)}\frac{v_S v_T}{3\Lambda}
\nn
\delta c_{23} &=&\frac{\sqrt6\abs{y}^2 x_B x_D}{M_2M_3}e^{-i\left(\phi_2+\phi_3\right)}\frac{v_S v_T}{3\Lambda}
\end{eqnarray}

Then for normal hierarchy, using $M_{1,2}\gg M_3$ gives
\begin{eqnarray}
X_{12} &\approx& 0\nn
X_{13} &\approx& \frac{i2 \sqrt{3}}{M_1} \left[\frac{x_B x_D v_S v_T}{3 \Lambda}e^{-i\left(\phi_1+\phi_3\right)}\right]^*\nn
X_{23} &\approx& -\frac{i \sqrt{6}}{M_1} \left[\frac{x_B x_D v_S v_T}{3 \Lambda}e^{-i\left(\phi_2+\phi_3\right)}\right]^*\nn
&\approx&-\frac{1}{ \sqrt 2}X_{13}
\end{eqnarray}
so that
\begin{eqnarray}
s_{13} &\approx& \sqrt{\frac16}\abs{X_{13}} =\abs{\frac{\sqrt2 x_B x_D v_S v_T}{3M_1 \Lambda}} \nn
\delta \theta_{12} &\approx&0\nn
\delta\theta_{23} &\approx& -\frac{2}{\sqrt 3} \text{Im}\left[e^{i\phi_{13}}X_{13}\right]
\end{eqnarray}
so that $\abs{\delta \theta_{23}} \le 2 \sqrt2  s_{13}$.

In conclusion, we have found that the models in the literature that generate an exactly tribimaximal PMNS mixing matrix using a flavor symmetry (in a small expansion parameter $\eta$) do not have leptogenesis. Higher order terms in $\eta$ can give an adequately large lepton asymmetry of order $\epsilon \sim \mathcal{O}\left(\eta^2/(32\pi)\right) \agt 10^{-6}$. These terms also lead to deviations from tribimaximal mixing, with $\theta_{13} \sim \mathcal{O}\left(\sqrt{32\pi \epsilon} \right) \sim \mathcal{O}\left(\eta\right) \agt 10^{-2}$ and correlated deviations from the tribimaximal values of $\theta_{23}$ and $\theta_{12}$ also of order $\eta$.  

There are also corrections to the light-neutrino mass matrix due to renormalization group running of $c_5$ between the seesaw scale and the weak scale. In particular, the Ma form Eq.~(\ref{tbform}) is not preserved by the $c_5$ anomalous dimension~\cite{run1,run2}. This induces a non-zero $\theta_{13}$ of order
\begin{eqnarray}
\theta_{13} \sim \frac{1}{16\pi^2} \left( \frac{m_\tau}{v} \right)^2 \log \frac{M}{v} \sim 10^{-5}
\end{eqnarray}
which is formally $\mathcal{O}\left(\eta^2\right)$ and much smaller than the effects we have considered.

\end{document}